\documentclass[twocolumn,preprintnumbers,amssymb,amsmath,superscriptaddress,nofootinbib]{revtex4}
\usepackage{graphicx}% Include figure files 
\usepackage{dcolumn}% Align table columns on decimal point 
\usepackage{bm}% bold math 
\usepackage{epsfig,amsmath,amssymb,verbatim,mathrsfs,array,layout,textcomp,amssymb,latexsym,color}
\usepackage{enumerate}
\input epsf.tex
\newcommand{\beq}{\begin{eqnarray}}% can be used as {equation} or {eqnarray}
\newcommand{\eeq}{\end{eqnarray}}

\def\ltap{\ \raise.3ex\hbox{$<$\kern-.75em\lower1ex\hbox{$\sim$}}\ }
\def\gtap{\ \raise.3ex\hbox{$>$\kern-.75em\lower1ex\hbox{$\sim$}}\ }

\def\eps{\epsilon}

\def\be{\begin{equation}}
\def\ee{\end{equation}}
\def\bea{\begin{eqnarray}}
\def\eea{\end{eqnarray}}
\newcommand{\eref}[1]{(\ref{#1})}
\newcommand{\Eref}[1]{Eq.~(\ref{#1})}
\newcommand{\Erefs}[1]{Eqs.~(\ref{#1})}

\newcommand{\BR}{{\rm BR}}
\newcommand{\abs}[1]{\left\lvert#1\right\rvert}
\newcommand{\cO}{\mathcal{O}}

\definecolor{red1}{cmyk}{0,1,1,0.3}

\begin{document} 

\title{Charming the Higgs}
\author{C\'edric Delaunay}
\affiliation{LAPTh, Universit\'e de Savoie, CNRS, B.P.110, F-74941 Annecy-le-Vieux, France}
\affiliation{CERN Physics Department, Theory Division, CH-1211 Geneva 23, Switzerland}
\author{Tobias Golling}
\affiliation{Department of Physics, Yale University, New Haven, CT 06520, USA}
\author{Gilad Perez}
\affiliation{CERN Physics Department, Theory Division, CH-1211 Geneva 23, Switzerland}
\affiliation{Department of Particle Physics and Astrophysics, Weizmann Institute of Science, Rehovot 76100, Israel}
\author{Yotam Soreq}
\affiliation{Department of Particle Physics and Astrophysics, Weizmann Institute of Science, Rehovot 76100, Israel}

\preprint{\scriptsize CERN-PH-TH/2013-252\vspace*{.1cm}}
\preprint{\scriptsize LAPTh-062/13}
\vskip .05in

\begin{abstract}
  \vskip .05in We show that current Higgs data permit a significantly
  enhanced Higgs coupling to charm pairs, comparable to the Higgs to
  bottom pairs coupling in the Standard Model, without resorting to
  additional new physics sources in Higgs production. With a mild level of the latter 
current data even allow for the Higgs to charm pairs to be the
 dominant decay channel.
 An immediate consequence of such a large
  charm coupling is a significant reduction of the Higgs signal
  strengths into the known final states as in particular into bottom
  pairs. This might reduce the visible vector-boson associated Higgs
  production rate to a level that could compromise the prospects of ever
  observing it.  We however demonstrate that a significant fraction of this reduced signal can be
 recovered by jet-flavor-tagging targeted towards
  charm-flavored jets. Finally we argue that an enhanced Higgs to charm pairs
   coupling can be obtained in various new physics scenarios in the presence of 
 only a mild accidental cancellation between various
  contributions.
\end{abstract}

\maketitle
%%%%%%%%%%%%%%%%%%%%%%%%%%%%%%%%%%%%%%%%%%%%%%%%%%%%%%%%%%%%%%%%%%%%%%%%%%%%%%

%%%%%%%%%%%%%%%%%%%%%%%%%%%%%%%%%%
\section{Introduction} \label{sec:intro}
%%%%%%%%%%%%%%%%%%%%%%%%%%%%%%%%%%

The recent discovery of a Higgs-like particle at the
LHC~\cite{Aad:2012tfa,Chatrchyan:2012ufa} is a remarkable success of
the Standard Model (SM) of particle physics. The current data imply
that the new particle is consistent with the SM
predictions~\cite{Falkowski:2013dza,Giardino:2013bma}. Still, a lot is
yet to be learned regarding the properties of this recently discovered
particle.

It is important to study the nature of the Higgs couplings to other SM
fields. As the Higgs is rather light, with a mass smaller than that of
the top quark and smaller than twice the mass of the $W$ and $Z$
boson, it decays to particles that very weakly interact with it. In
fact, the dominant decay mode of the Higgs is to a bottom pair within
the SM, and the bottom Yukawa coupling to an on-shell Higgs is
$\mathcal{O}(0.02)$.  This exposes the Higgs branching ratios to a
generic susceptibility to any form of new physics. Any deformation of
the Higgs couplings to the SM particles, or introduction of additional
couplings to new fields, that competes with the small Higgs to bottom
coupling will lead to a significant change of the Higgs phenomenology at
the LHC. An interesting picture emerges from the potential changes to
the existing Higgs couplings to SM particles.  The next to leading
five couplings beyond the bottom coupling are Higgs couplings to $W$,
$Z$ and $\tau$, which are already measured to decent accuracy, the
coupling to gluons which controls the Higgs production cross section,
and the coupling to the charm quark. Among those SM states the charm
stands out as almost nothing is known experimentally on its 
coupling to the Higgs boson.

The Higgs branching ratio into charm pairs is $\mathcal{O}(3\%)$ in
the SM, which renders any attempt to directly probe the Higgs to charm
coupling at the LHC extremely challenging due to the large multijet
background. However, the charm Yukawa coupling is only few times
smaller than the bottom one in the SM, about one part in five at the Higgs mass scale~\cite{Xing}. Thus, only a factor of few 
mismatch between the actual charm coupling and its SM value would lead
to a significant change of the Higgs phenomenology. Furthermore, despite
its small value, the charm mass is not negligible and
due to the CKM suppression in bottom decays
 the charm
 life time is comparable to that of the bottom quark. Hence, jets originating from charm quarks
can in principle be identified at colliders.  The ATLAS collaboration,
in fact, recently presented new experimental techniques designed to
tag charm jets at the LHC~\cite{ATLAS2013068}. The possibility to
apply charm-tagging, beyond its plain interest from the SM
perspective~\cite{Chatrchyan:2013uja, Bodwin:2013gca}, also opens new
possibilities to analyze various beyond the SM
signals~\cite{Mahbubani:2012qq,Blanke:2013uia,DaRold:2012sz, Delaunay-Flacke-Gonzalez-Fraile-Lee-Panico-Perez}. In particular we
show in this paper how crucial charm-tagging may be in order to exhume
the associated Higgs production signal in the case of a suppressed
$h\to b\bar{b}$ branching ratio due to an enhanced Higgs to charm
coupling relative to the SM.

There is currently no attempt to directly probe the $h\to c\bar{c}$
channel at colliders and the Higgs to charm coupling is constrained
indirectly through the bound on the allowed Higgs ``invisible'' (more
precisely, unobserved) branching ratio. For SM Higgs production cross
sections this branching ratio cannot exceed $\sim 20\%$ at 95\%~
confidence level (CL), or $\sim 50\%$ if an additional new physics
source of gluon fusion production is
assumed~\cite{Falkowski:2013dza}. This implies a rough upper bound on
the Higgs to charm coupling of about three to five times its SM value,
assuming no other source of invisible decays other than Higgs decays
into the unobserved SM states. Hence, the current data still allow
for the $h\to c\bar c$ decay channel to be comparable in size with or even to dominate over  the $h\to b\bar b$ one.

The outline of the paper is as follows. In the next section we provide
a quantitative analysis of the current Higgs data in order to derive
the present bounds on the Higgs to charm coupling. We then demonstrate
in Section~\ref{sec:VH} that an enhanced charm coupling significantly
suppresses the $h\to b\bar b$ signal strength in associated Higgs
production, mostly through a reduced Higgs branching ratio into bottom pairs, and that the SM level of this signal could be partially or even entirely recovered by enriching the sample with charm-tagged
events, depending on the charm-tagging efficiency. In Section~\ref{sec:BSM}, we argue that a large Higgs to charm
coupling can be obtained under reasonable conditions in various
theories beyond the SM where moderate cancellation is present. We
present our conclusions in Section~\ref{sec:Summary}.

%%%%%%%%%%%%%%%%%%%%%%%%%%%%%%%%%%
\section{Constraints from Higgs data}  \label{sec:PrimilFit}
%%%%%%%%%%%%%%%%%%%%%%%%%%%%%%%%%%

A Higgs to charm pair coupling significantly larger than in the SM affects
both Higgs production cross sections and branching ratios, and is
therefore indirectly contrained by current Higgs rate measurements at
the LHC. On the one hand a large Higgs to charm coupling implies a
universal reduction of all Higgs branching ratios other than into
$c\bar c$ final states, provided all other Higgs couplings remain
standard. On the other hand Higgs production at hadron colliders is
also typically enhanced relative to the SM through a more important
charm fusion mechanism occurring at tree-level. (Another effect,
though far subdominant, arises in gluon fusion Higgs production
through a modified charm-loop contribution.) Therefore, one might
expect that there is a charm coupling value for which the enhancement in
Higgs production approximately compensates the universal suppression
in Higgs decays so that Higgs rates measured at the LHC remain close
to the SM predictions. We thus perform a fit of all available Higgs
data allowing deviations of the $hc\bar c$ coupling relative to the
SM in order to quantitatively determine the largest value presently
allowed.

We follow the approach of Ref.~\cite{Falkowski:2013dza} to globally fit available Higgs data. We consider both direct data from Higgs rate measurements at the LHC and indirect constraints from electroweak (EW) precision measurements at LEP. We assume that there is only one Higgs scalar $h$ of mass  $m_h=126\,$GeV, which is a singlet of the custodial symmetry preserved by EW symmetry breaking (EWSB). The Higgs interactions with other SM particles are assumed to be flavor-conserving and accurately enough parameterized by the effective Lagrangian
\beq\label{Leff}
	\mathcal{L}_{\rm eff} = \mathcal{L}_{0} + \mathcal{L}_{2}\,,
\eeq
where interactions to zeroth-order in derivatives are 
\beq
\mathcal{L}_{0} &=& \frac{h}{v}\Big[  c_V \left( 2m_W^2 W^+_\mu W^{\mu-} + m_Z^2 Z_\mu Z^\mu \right) \nonumber\\
&&\quad\quad
	 - \sum_{q} c_q m_q \bar{q} q - \sum_\ell c_\ell m_\ell \bar \ell \ell	\Big]\,,
\eeq
and interactions to next-to-leading order in derivatives are
\beq
\mathcal{L}_{2}& =&\frac{h}{4v}\Big[c_{gg} G^a_{\mu\nu} G^{\mu\nu a}- c_{\gamma\gamma} F_{\mu\nu} F^{\mu\nu}- 2c_{WW} W^+_{\mu\nu} W^{\mu\nu-}  \nonumber\\
&&\quad\quad\quad - 2c_{Z\gamma}F_{\mu\nu}Z^{\mu\nu}
	- c_{ZZ} Z_{\mu\nu} Z^{\mu\nu} 	 \Big] \, ,
\eeq
where $q=u,d,s,c,b,t$ and $\ell=e,\mu,\tau$ are the SM massive quarks and charged leptons, $v=246\,$GeV is the EWSB scale, $W_\mu$, $Z_\mu$, $A_\mu$ and $G_\mu$ are the SM gauge fields with the corresponding fields strength tensors. The tree-level SM limit is achieved by  $c_V=c_q=c_\ell=1$ and $c_{\gamma\gamma}=c_{gg}=c_{Z\gamma}=0$ (before the top quark has been integrated out).
We neglect CP-odd operators and assume real $c_{q,\ell}$ coefficients as there is only a weak sensitivity to CP-odd couplings and CP-violating phases in Higgs rate measurements.\footnote{Higher sensitivities to CP-odd couplings may be reached for instance through angular distribution measurements, in particular in vector-boson associated Higgs production channels where the Higgs boson can be significantly boosted~\cite{CPinVH,Hupdown}.} (See {\it e.g.}  Ref.~\cite{Harnik-Zupan} for a recent update.)  
The underlying custodial symmetry imposes the following relations among couplings in $\mathcal{L}_{2}$~\cite{Falkowski:2013dza}
\beq
c_{WW} =c_{\gamma\gamma} + \frac{g_L}{g_Y} c_{Z\gamma}\,,\quad c_{ZZ} = c_{\gamma\gamma} + \frac{g_L^2 - g_Y^2}{g_Y g_L} c_{Z\gamma}\,,
\eeq
where $g_L$ and $g_Y$ are the SU(2)$_L$ and U(1)$_Y$ gauge couplings, respectively.
In contrast with existing Higgs fits, as in {\it e.g.} Refs.~\cite{Carmi:2012in,Falkowski:2013dza,Giardino:2013bma,Belanger:2013xza}, we leave the $c_c$ as a free parameter of the fit. Current Higgs data are very unlikely to be sensitive to Higgs couplings to $e$, $\mu$, and $u$, $d$, $s$, as the latter are already very small in the SM. We thus set $c_{e,\mu}=c_{u,d,s}=1$ in the following. We are left with at most eight independent free parameters: $c_V$, $c_{c,b,t}$, $c_\tau$, $c_{gg}$, $c_{\gamma\gamma}$ and $c_{Z\gamma}$. 

The Higgs rate measurements at the LHC are presented in the form of signal strengths defined as
\beq\label{eqmu}
	\mu_f \equiv \frac{\sigma_{pp\to h}\, \BR_{h\to f }}{\sigma_{pp\to h}^{\rm SM}\, \BR_{h\to f} 
^{\rm SM}} \, ,
\eeq
for each final state $f$, where $\sigma_{pp\to h}$ and $\BR_{h\to f}$ are the Higgs production cross section and branching ratio, respectively, while the $^{\rm SM}$ label denotes their corresponding SM predictions. Similar signal strengths measured at the Tevatron are obtained from \Eref{eqmu} through the replacement $pp\to p\bar p$. 
We perform a standard $\chi^2$ analysis in order to fit the coefficients in \Eref{Leff}  to current Higgs data. The total $\chi^2$ function is 
\beq\label{chi2}
	\chi^2 = \sum_{f,i} \frac{\left(\mu_{f,i}^{\rm th} - \mu_{f,i}^{\rm ex}\right)^2}{\sigma_{f,i}^2}  \, ,
\eeq
where the index $i$ runs over all measurements of the channel $f$ and correlations between different channels are neglected.  
$\mu_{f,i}^{\rm ex}$  and $\sigma_{f,i}$ denote the experimental central values and their corresponding standard deviations, respectively. Asymmetric experimental errors are symmetrized  for simplicity.  We consider the most updated set of Higgs measurements in $h\to WW^*,ZZ^*$ and $\gamma\gamma$ channels from ATLAS~\cite{ATLASdata}, CMS~\cite{CMSdata} and Tevatron~\cite{Aaltonen:2013kxa} collaborations, as well as the $h\to \tau\tau$ results from CMS~\cite{CMS_tautau} and Tevatron~\cite{Aaltonen:2013kxa}. We also include the recent $h\to b\bar b$ search in vector-boson associated production~\cite{CMS_VHbb}
 and in vector-boson fusion at CMS~\cite{CMS_VBFbb}, as well as the $h\to Z\gamma$ search at CMS~\cite{CMS_Zphoton}. We do not use the recent $h\to b\bar b$ and $h\to \tau \tau$  preliminary ATLAS results. However, we checked that the latter does not significantly change  our results given the current experimental sensitivity in these channels.
$\mu_{f,i}^{\rm th}$ are the theoretical signal strength predictions, which incorporate the relative weights of each Higgs production mechanisms as quoted by the experimental collaborations, whenever available. This is the case for all channels that we use except for $Vh(b\bar b)$ at CMS for which we assume pure vector-boson associated production.
Theoretical predictions for Higgs signal strengths in terms of the effective coefficients in \Eref{Leff} can be found in Ref.~\cite{Falkowski:2013dza}, while we use the SM Higgs production cross sections and branching ratios of Ref.~\cite{LHCHiggsCrossSectionWorkingGroup:2011ti}. We however add the following two modifications in order to implement a $hc\bar c$ coupling significantly different than its SM value. First of all, we include the charm loop contribution in the gluon fusion cross section as $\sigma_{gg\to h}/\sigma_{gg\to h}^{\rm SM}\simeq \abs{\hat c_{gg}}^2/\abs{\hat c_{gg}^{\rm SM}}^2$ with 
\beq\label{hatcgg}
\hat c_{gg} &=& c_{gg}+\Big[1.3\times 10^{-2} c_t-\left(4.0-4.3 i\right)\times 10^{-4}c_b\quad\quad\nonumber\\
&&\quad\quad\quad\quad\quad\quad -\left(4.4-3.0 i\right)\times 10^{-5}c_c\Big]\,,
\eeq
 where numbers are obtained using the running quark masses extracted from Ref.~\cite{Xing}. $\hat c_{gg}^{\rm SM}\simeq 0.012\,$ is obtained by taking the SM limit, $c_{gg}\to 0$ and $c_{t,b,c}\to 1$, in \Eref{hatcgg}. Then, we include the charm fusion cross section as  $\sigma_{c\bar c\to h}\simeq 3.0\times 10^{-3}\, |c_c|^2\,\sigma_{gg\to h}^{\rm SM}$, where the charm fusion to gluon fusion cross section ratio is evaluated at next-to-leading order in the QCD coupling and we use MSTW parton distribution functions~\cite{Martin:2009iq}. We transposed the NLO bottom fusion cross section obtained in Ref.~\cite{Harlander:2003ai} in order to estimate $\sigma_{c\bar c\to h}^{\rm SM}$.\\

\begin{figure}[!t]
  \begin{center}
  \includegraphics[width=.5\textwidth]{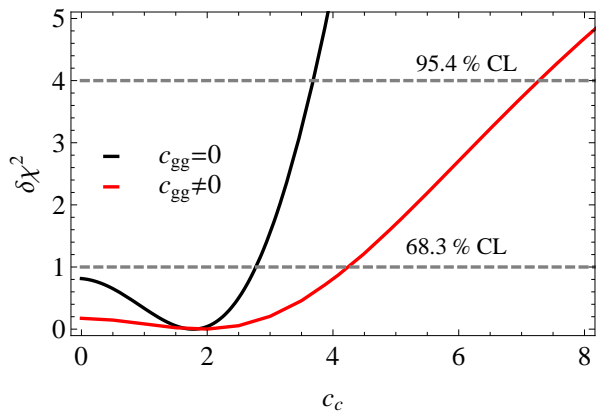}
  \caption{$\delta\chi^2=\chi^2-\chi^2_{\rm min}$ as a function of the Higgs to charm pairs coupling $c_c$. The black and red curves correspond respectively to  case (a), where all Higgs coupling but $c_c$ are SM-like, and case (b), where only $c_c$ and $c_{gg}$ deviate from the SM and marginalizing over the latter. Horizontal dashed lines denotes the $68.3\%$ and $95.4\%$ CL ($\delta\chi^2=1$ and $4$, respectively).}
  \label{fig:dchi2charm}
  \end{center}
\end{figure}
\begin{figure}[!t]
  \begin{center}
  \includegraphics[width=.45\textwidth]{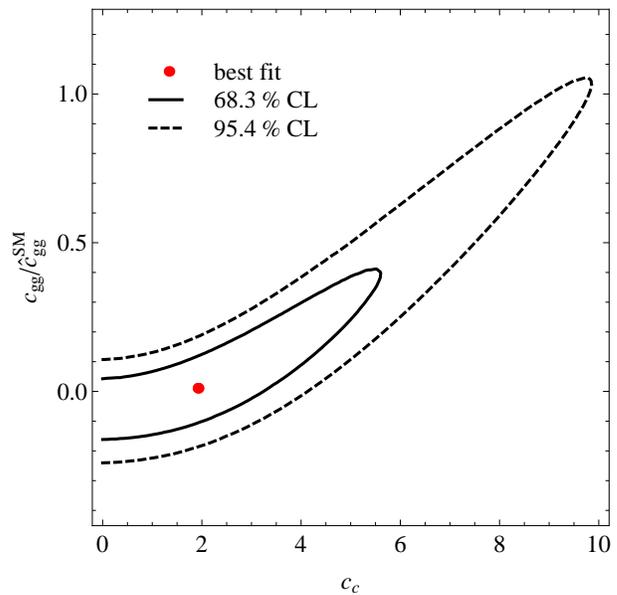}
  \caption{$68.3\%$ (solid) and $95.4\%$ (dashed) CL regions in the
    $c_c-c_{gg}$ plane in case (b) where only the Higgs to charm and
    Higgs to gluon couplings are allowed to deviate from their SM
    values. The red dot represents the best fit point. $\hat
    c_{gg}^{\rm SM}\simeq 0.012\,$.
  }
  \label{fig:charmggCLs}
  \end{center}
\end{figure}

We mainly focus on two different scenarios, where 
\begin{enumerate}[(a)]
\item all Higgs couplings but the charm one are SM-like.
\item all the Higgs couplings but the charm one and $c_{gg}$ are SM-like.
\end{enumerate} 
The general case, where all independent parameters are allowed to deviate from the SM is discussed in Appendix~\ref{append:gen}. However, the results are found to be very close to those from case (b) above. Thus, unless explicitly mentioned otherwise, case (b) can be taken as a proxy for the general case. In both cases (a) and (b), $\chi^2$ only depends on the Higgs to charm coupling through $\abs{c_c}^2$, up to a small interference effect with top and bottom loops in gluon fusion production. Hence, there is almost no sensitivity to the sign of $c_c$. For simplicity we consider positive $c_c$ in the following.
$\chi^2$ minimization yields 
\begin{align} \label{eq:cc1}
	c_c \leq 3.7\ (7.3)\,,\quad {\rm at}\ 95.4\%\ {\rm CL}\,,
\end{align}
for case (a) (case (b)) as defined above. The bound in case (b) is obtained upon marginalizing over the $c_{gg}$ coupling. The larger allowed  range for the charm coupling in case (b), relative to case (a), is due to a further enhancement of the Higgs production cross section from $c_{gg}>0$. $\delta\chi^2\equiv\chi^2-\chi^2_{\rm min}$ as a function of $c_c$ for both cases is shown in Fig.~\ref{fig:dchi2charm}. The $\delta\chi^2$ raise in case (b) for $c_c\gtrsim 3$ is due to the further universal suppression in the branching ratios induced by the larger $c_{gg}>0$ values required to compensate for the increase in $c_c$. For case (b), we also show in Fig.~\ref{fig:charmggCLs} the $68.3\%$ and $95.4\%$ CL regions in the $c_c-c_{gg}$ plane.
$h\to WW^*$ is the most significant channel which dominantly drives
the $\chi^2$
fit. 
Since $\mu_{WW^*}\lesssim1$ at both ATLAS and CMS experiments, a total
Higgs width slightly larger than in the SM is favored. This results in
the fact that the $\chi^2$ takes a minimum at a larger charm coupling,
$c_c\gtrsim1$, as shown in Fig.~\ref{fig:dchi2charm}. Excluding the
$h\to WW^*$ channel, the remaining average signal strength becomes
$\gtrsim 1$ and the $\chi^2$ fit favors lower values of $c_c$. \\

We conclude that without additional new physics contributions to Higgs
production other than the contribution to charm and gluon fusion, the
latter being subdominant, a Higgs to charm coupling as large as about four times
its SM value, is consistent with current Higgs data within $95.4$\%
CL. Even larger charm couplings are allowed at $95.4\%$ CL,
provided that there is a conjoint $\cO(1)$ enhancement in gluon fusion
production from a new physics source. Such a large $hc\bar c$ 
coupling would in particular significantly reduce the Higgs branching ratio in bottom pairs. Consequently, suppressed $h\to b\bar b$ signals in
vector-boson associated Higgs production at ATLAS and CMS experiments are expected as these channels are much less sensitive to gluon fusion and $c\bar c$ fusion production mechanisms.
Needless to say even bigger
effects are found when the other couplings are allowed to float as
well, in particular the higgs to gluons effective coupling.

%%%%%%%%%%%%%%%%%%%%%%%%%%%%%%%%%%
\section{Observability of $h\to c\bar{c}$ at the LHC} \label{sec:VH}
%%%%%%%%%%%%%%%%%%%%%%%%%%%%%%%%%%

We showed in the previous section that a Higgs coupling to $c\bar c$
significantly larger than in the SM is allowed by current Higgs
data. We argue here that such a large coupling yields important
effects in channels where the Higgs boson decays into bottom pairs. In
particular, one expects a significant suppression of $\mu_{b\bar b}$
in vector-boson associated prodution, due to a sizable enhancement of
$\BR_{h\to c\bar c}$ relative to the SM. We also demonstrate that the
associated production signal can be partially recovered by using the
recently developed charm-tagging technique~\cite{ATLAS2013068}.\\

We identify the following three interesting phenomenological aspects
of having a large Higgs to charm  coupling.  First of all, $b\bar b$
signal strengths in associated production are suppressed due to the
larger Higgs width which reduces the branching ratio into bottom pairs
as\footnote{We neglected in \Eref{BRbb} the subdominant effect of the Higgs to charm coupling in the loop induced $h\to gg$ partial width.}
\beq\label{BRbb}
\frac{\BR_{h\to b\bar b}}{\BR_{h\to b\bar b}^{\rm SM}}&=&\Bigg[1+(\abs{c_c}^2-1)\BR_{h\to c\bar c}^{\rm SM}\nonumber\\
&&+\left(\abs{\frac{c_{gg}}{\hat c_{gg}^{\rm SM}}+1}^2-1\right)\BR_{h\to gg}^{\rm SM}\Bigg]^{-1}\,.
\eeq
\Erefs{eq:cc1} and \eref{BRbb} show that enhancing the $hc\bar c$ coupling results in a significant reduction in the Higgs to bottom pairs rate in associated production processes $Vh(\bar b b)$ of
\beq
\mu_{b\bar b} \simeq 0.74 \, (0.40)\,,
\label{mubb}
\eeq
for case (a) ((b)) respectively, where we assumed SM-like $Vh$ production and no acceptance for the other production mechanisms. The suppressed signal in \Eref{mubb} is still consistent at $95.4\%$ CL with all other existing Higgs data. This result  
makes this final state extremely challenging for the next run of the LHC. 
Note that in case (b) we include the subdominant Higgs width increase coming from $c_{gg}>0$, which further suppresses $\BR_{h\to b\bar b}$.

\begin{figure}[!t]
  \begin{center}
    \includegraphics[width=.45\textwidth]{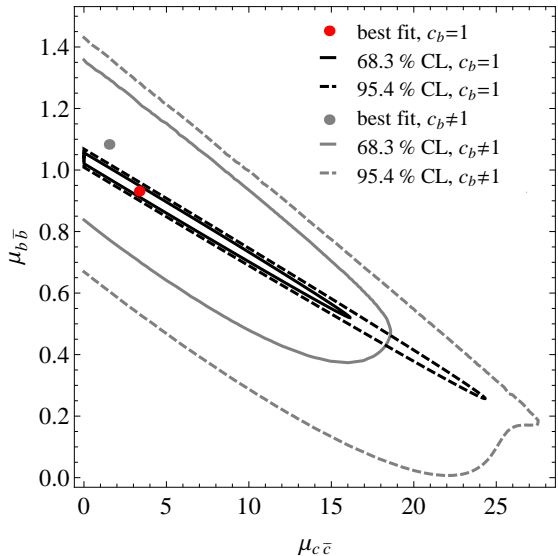}% \hspace{1cm}
\end{center}   
 \caption{Correlation between $\mu_{c\bar{c}}$ and $\mu_{b\bar{b}}$ signal strengths in the presence of an enhance Higgs to charm coupling relative the SM. Relative weights of each Higgs production mechanisms from the CMS analysis~\cite{CMS_VHbb} are assumed for both signal strengths. The red dot represents the best fit point and the solid (dashed) black line is the $68.3\, (95.4)\%$~CL contour derived from fitting current Higgs data for case (b), where only $c_c$ and $c_{gg}$ are not SM-like. The solid (dashed) gray contour delineates the allowed $68.3\,(95.4)\%$~CL region for the more general case where $c_b$ is allowed to vary as well.}
  \label{fig:muccVSmubb1}  
\end{figure}

Second, we stress that there is an correlation between the
measured $\mu_{b\bar{b}}$ and $\mu_{c\bar{c}}$ signal strengths, as
the production cross section is identical for both channels and increasing
the Higgs to charm coupling ($c_c>1$) leads to a suppressed branching
ratio into $b\bar b$ and an enhanced one into $c\bar c$. More
precisely the signal strengths into bottom and charm pairs
are simply proportional to each other, $\mu_{c\bar c}=
|c_c/c_b|^2\mu_{b\bar b}\,$. Deviations of the bottom coupling from the SM limit are much more constrained by Higgs data than for the charm coupling (see Eq.~\eqref{fullfit}), which yields $\mu_{c\bar c}\simeq |c_c|^2\mu_{b\bar b}$.  
Moreover, we find that this strong correlation
remains also in the presence of an additional new physics source of
gluon fusion production ($c_{gg}\neq0$), as illustrated in
Fig.~\ref{fig:muccVSmubb1} which shows the regions of the
$\mu_{b\bar{b}}-\mu_{c\bar{c}}$ plane consistent within $68.3\%$ and
$95.4\%$ CL with Higgs data. The signal strengths in
Fig.~\ref{fig:muccVSmubb1} are evaluated assuming the relative weights of Higgs production mechanisms used by the CMS $b\bar{b}$
search in associated Higgs production~\cite{CMS_VHbb}.
Figure~\ref{fig:muccVSmubb1} also shows that in the case that $c_{c}$,
$c_{gg}$ and $c_b$ are all allowed to vary, the above correlation
still endures, albeit in a weaker way.

Finally, we find that the expected combined signal strength into bottom and charm pairs can be relatively enhanced compared to that of only bottom pairs, despite the smaller charm-tagging efficiency. We define the combined signal strength into  $b\bar b$ and $c\bar c$ as
\beq\label{combined}
	\mu_{b\bar b+c\bar c} \equiv
	\frac{\sigma_{pp\to h}\left( \varepsilon_b^2\BR_{h\to b\bar b} + \varepsilon_c^2\BR_{h\to c\bar c} \right)}
	{\sigma_{pp\to h}^{\rm SM}\left( \varepsilon_b^2\BR_{h\to b\bar b}^{\rm SM} + \varepsilon_c^2\BR_{h\to c\bar c}^{\rm SM} \right)} \,, 
\eeq 
where $\varepsilon_{b}$ and $\varepsilon_c$ are the tagging efficiencies for bottom and charm jets, respectively. We implicitly assumed in \Eref{combined} that the only difference between $c\bar c$ and $b\bar b$ analyses is the tagging efficiency, and that, in particular, the production cross section is the same in both cases. 
We also define the ratio of the combined and bottom only signal strengths as
\beq
	R \equiv \frac{\mu_{b\bar{b}+c\bar{c}}}{\mu_{b\bar{b}}}
= 	\frac{1 + \abs{c_c}^2r_{cb}^2\,\BR_{h\to c\bar c}^{\rm SM}/\BR_{h\to b\bar b}^{\rm SM}}{1 +r_{cb}^2\,\BR_{h\to c\bar c}^{\rm SM}/\BR_{h\to b\bar b}^{\rm SM}} \, ,
\eeq
where $r_{cb}\equiv\varepsilon_c/\varepsilon_b$ is the ratio of tagging efficiencies of charm- and bottom-flavored jets. Assuming the branching ratio values for a 126$\,$GeV SM Higgs~\cite{LHCHiggsCrossSectionWorkingGroup:2011ti}, the upper bound from \Eref{eq:cc1} derived from fitting current Higgs data implies  
\beq\label{R}
R\, \lesssim \, 1+0.21 \, (0.86) \,\times \,\left(\frac{r_{cb}}{0.57}\right)^2\,,
\eeq
%TG: why is R always larger than 1?  For epsilon_c= 0 shouldn't we get back Eq. (10)?
%CDTGYS: add Eq. that multiply \mu_bb * R and explain it. 
for the case (a) ((b)) defined above, where $r_{cb}\simeq 0.57$ corresponds to $\varepsilon_b\simeq0.7$~\cite{CMS_VHbb} together with a prospective charm-tagging efficiency of $\varepsilon_c\simeq 0.4$. The parameter $R$ only measures how much combining charm and bottom pairs enhances the associated production signal relative to a sample of bottom pairs only. In particular it is independent of the production cross section by construction. The enhancement in \Eref{R} is to be compared with the much reduced signal strength available when only $b$-tagging is used to extract the signal, as given in Eq.~\eqref{mubb}. Combining Eqs.~\eqref{mubb} and~\eqref{R} yields a combined signal strength of 
\beq
\mu_{b\bar b+c\bar c}= 0.89\, (0.75)\,,
\eeq
for case (a) ((b)), where $\varepsilon_c\simeq 0.4$ as well as pure SM-like vector-boson associated production are assumed.\\

We showed that the expected bottom pair signal strength in associated Higgs
production can be significantly reduced, relative to the SM, in the
presence of a largely enhanced $hc\bar c$ coupling, since the Higgs
decay to charm pairs is becoming as important. Moreover, \Eref{R}
shows that an increased tagging efficiency for charm jets can bring
the measured associated Higgs production signal strength almost back to its SM
level.

%%%%%%%%%%%%%%%%%%%%%%%%%%%%%%%%%%
\section{Charm coupling beyond the SM} \label{sec:BSM}
%%%%%%%%%%%%%%%%%%%%%%%%%%%%%%%%%%

Modified Higgs couplings to fermions can arise in many  theories beyond the SM. We consider here different theoretical frameworks which illustrate the possibility of having a Higgs to charm  coupling significantly larger than within the SM. We begin with an effective field theory (EFT) discussion, where new physics above the weak scale is described by a set of higher-dimensional operators, in order to stress that Higgs coupling to charm is {\it a priori} not necessarily related to the small charm mass. We then discuss how much the Higgs to charm coupling can deviate from the SM within specific new physics scenarios. In particular, we show that it is possible to obtain $c_c$ of order few, with all other couplings SM-like, in a two Higgs doublet model (2HDM) with minimal flavor violation (MFV) and in a General MFV (GMFV)~\cite{Kagan:2009bn} scenario with only one Higgs doublet. We finally comment on composite models where the Higgs field is realized as a pseudo-Nambu--Goldstone boson (pNGB).\\

Within the EFT framework the Higgs to charm coupling is modified in the presence of a dimension six operator. The relevant operators in the up-type quark sector are
\beq\label{LEFT}
	\mathcal{L}_{\rm EFT} 
\supset	\lambda^u_{ij}  \bar{Q}_{i}\tilde H  U_{j} 
	 +\frac{g^u_{ij}}{\Lambda^2}  \bar{Q}_{i}\tilde H U_{j} \left( H^\dagger H \right)  + {\rm h.c.} \, ,
\eeq
where the first term is the marginal up-type Yukawa operator of the SM and the second term is a dimension six  operator suppressed by the new physics scale $\Lambda$. $Q_i$ and $U_i$, with $i=1,2,3$, are the SM quark left-handed doublets  and right-handed singlets, respectively, and  $H$ is the Higgs doublet, with $\tilde{H}=i\sigma_2H^*$. $\lambda^u$ and $g^u$ are generic complex $3\times3$ matrices in flavor space. Setting the Higgs field to its vacuum expectation value $H=(0,(v+h)/\sqrt{2})^T$, the mass and linear Higgs coupling matrices are respectively
\beq
	M^u_{ij} 
&=&	\frac{v}{\sqrt{2}}\left( \lambda^u_{ij} +  g^u_{ij}\frac{v^2}{2\Lambda^2} \right)\,, \\
	Y^u_{ij} 
&=&	\frac{1}{\sqrt{2}}\left( \lambda^u_{ij} + 3 g^u_{ij}\frac{v^2}{2\Lambda^2} \right)\,.
\eeq
We assume for convenience that $\lambda^u$ and $g^u$ are aligned and that only $g^u_{22}\neq0$ in the mass basis. In this case the deviation from the SM Higgs to charm coupling is simply 
\beq
c_c 
= 	1 + \frac{3}{2}\frac{v^2}{\Lambda^2} \frac{g^u_{22}}{ y_c} \, ,
\eeq
where we defined $y_c \equiv \sqrt{2}\, m_c/v\simeq 3.6\times 10^{-3}$, $m_c$ being the running charm quark at the Higgs mass scale~\cite{Xing}. Naive dimensional analysis suggests that the effective description breaks down at the scale $\Lambda$ for $g^u_{22}\sim16\pi^2$. As a function of the Higgs to charm coupling modification this scale is
\beq
	\Lambda \simeq \frac{63\, {\rm TeV}}{\sqrt{|c_c -1| }} \,.
\eeq
Assuming the upper bound on $c_c$ in \Eref{eq:cc1}, we find that the  cutoff scale can be as high as  $\Lambda \lesssim 38\,(25)\,$TeV for case (a) ((b)). These scales are sufficiently high so that it is possible that the associated new physics  dynamics at the cut-off leaves no direct signatures at the LHC other than a significantly enhanced Higgs to charm coupling. \\

We now focus on some specific new physics scenarios. Consider a 2HDM with MFV~\cite{Jung:2010ik, Trott:2010iz}. 
In this setup, the MFV ansatz allows to write the SM-like Higgs couplings to fermions as an expansion in the spurionic parameters which break the flavor symmetry group. Following notations of Ref.~\cite{Dery:2013aba}, the charm and top quark couplings to the SM-like Higgs boson are  
\beq	c_t 
&\simeq&	 A^U_S + B^U_S y^2_t + C^U_S y^2_b \abs{V_{tb}}^2  \,, \nonumber\\
\label{c2HDM}	c_c 
&\simeq&	 A^U_S + B^U_S y^2_c + C^U_S \left( y^2_b \abs{V_{cb}}^2 + y^2_s\abs{V_{cs}}^2 \right) \,,
\eeq
where $y_i \equiv \sqrt{2}\, m_i/v$, $V_{ij}$ are the CKM matrix elements and $A^U_S$, $B^U_S$ and $C^U_S$ are $\mathcal{O}(1)$ coefficients. $\mathcal{O}(y_i^4)$ and higher contributions were neglected in \Eref{c2HDM}. Assuming for instance $A^U_S\simeq4$ and $B^U_S\simeq-3$, \Eref{c2HDM} yields $c_c\simeq4$ and $c_t\simeq1$. Moreover, in the limit where all the heavier Higgs states are decoupled, $c_V\simeq1$~\cite{Gunion:2002zf}. Therefore, a significantly larger charm coupling, with all other couplings close to their SM values,  can be obtained at the expense of a mild cancellation, at the level of one part in few, among unknown $\mathcal{O}(1)$ coefficients.\\

Consider now a model with one Higgs doublet in the GMFV framework~\cite{Kagan:2009bn}, in which large top Yukawa effects are resummed to all orders. We define our notations in Appendix~\ref{append:gmfv}.
In the mass basis, the up-type quark mass and linear Higgs interaction matrices become 
$M^u \simeq\lambda v/\sqrt 2\times$diag$\big(y_u(\gamma+\zeta x),\, y_c(\gamma+\zeta x),\, 1+ r x\big)$
and  to leading order in $\lambda_C\simeq0.23$, the sine of the Cabibbo angle and in $x\equiv v^2/(2\Lambda^2)\,$, with $\Lambda$ the GMFV scale, we find
\begin{align}
&	Y^u \simeq \frac{\lambda}{\sqrt{2}} \label{eq:GMFVYu}
	\begin{pmatrix}
		y_u(\gamma+3\zeta x) & 0 & 2\lambda_C^3 (\kappa - \alpha r) x\\
		0 & y_c(\gamma+3\zeta x) & 2\lambda_C^2 ( \kappa - \alpha r)x \\
		2y_u\lambda_C^3w x  & 2y_c\lambda_C^2w x & 1+ 3r x
	\end{pmatrix}\,,
\end{align}
where $w\equiv \eta - \gamma r + \alpha^*(\zeta - \gamma r)$.
Equation~\eqref{eq:GMFVYu} yields the following Higgs to charm coupling ratio in GMFV and in the SM 
\begin{align}
	c_c = \lambda(\gamma+3 \zeta x )  \simeq 1 + 2\lambda \zeta x \,.
\end{align} 
As $\lambda,\zeta\sim \mathcal{O}(1)$ and $x\lesssim 1$, $c_c>1$ can be obtained for not too small value of $x$. As in all the above cases the coupling enhancement is at the cost of a moderate accidental cancelation among $\mathcal{O}(1)$ couplings. 
Note that the GMFV scale is constrained through the off-diagonal entries in \Eref{eq:GMFVYu} by a series of flavor changing observables analysed in Ref.~\cite{Harnik-Zupan}. However, constraints from single-top production, neutral $D$ meson mixing, flavor changing top decay $t\to hj$ and neutron electric dipole moment (assuming $\mathcal{O}(1)$ phases in the fundamental parameters) are satisfied for $x\lesssim 1$ since GMFV contributions are suppressed by $\lambda_C^2$, $\lambda_C^5$, $\lambda_C^2$ and $y_u\lambda_C^6$, respectively.

Consider finally composite pNGB Higgs models. Modifications of Higgs couplings to up-type quark in composite Higgs models is parameterized by the effective Lagrangian in \Eref{LEFT} with $\Lambda$ remplaced by the global symmetry breaking scale $f$, the ``decay constant'' of the pNGB Higgs~\cite{Giudice:2007fh}. The dimension six coefficient in \Eref{LEFT} receives two types of contributions from the composite dynamics, $g^u=g^u_h+g^u_\psi$. The first term is a direct contribution from the non-linear Higgs dynamics and it is aligned with the marginal operator $g_h^u\propto \lambda^u$. The second term arises from the presence of light fermionic resonances from the strong dynamics. It is generically misaligned with $\lambda^u$ and its entries scale like $g^u_\psi\sim \lambda^u \eps^2(g_\psi f/m_\psi)^2$, where $g_\psi<4\pi$ and $m_\psi$ are respectively a typical strong coupling and a resonance mass of the strong dynamics, and $\eps<1$ is the degree of the compositeness of the SM quarks. Neglecting flavor violation for simplicity and assuming relatively composite right-handed charm quark, the Higgs to charm coupling is~\cite{Delaunay:2013iia}
\beq
c_c \simeq 1+ \mathcal{O}\left(\frac{v^2}{f^2}\right)+\mathcal{O}\left(\epsilon_c^2\frac{g_\psi^2v^2}{m_\psi^2}\right)\,,
\eeq
where $\eps_c$ is the right-handed charm degree of compositeness. The symmetry breaking scale $f$ is constrained by EW precision parameters to be $f\gtrsim 750\,$GeV (see {\it e.g.}~\cite{Grojean:2013qca,Ciuchini:2013pca} for a recent analysis). Hence, in the absence of light composite resonances associated with the charm quark, the Higgs to charm coupling is not expected to deviate significantly from its SM value. However, if light charm partner resonances are present a larger Higgs to charm can be obtained. Current bounds on the charm partner mass from direct searches  at the LHC are  $m_\psi\gtrsim \mathcal{O}(500\,$GeV$)$~\cite{Delaunay-Flacke-Gonzalez-Fraile-Lee-Panico-Perez}. Hence, for a fully composite charm quark $\eps_c\simeq 1$, $g_\psi \sim4\pi$ a largely enhanced $hc\bar c$ coupling is possible.

%%%%%%%%%%%%%%%%%%%%%%%%%%%%%%%%%%%
\section{Conclusions} \label{sec:Summary}
%%%%%%%%%%%%%%%%%%%%%%%%%%%%%%%%%%%

We pointed out that the Higgs to charm coupling can be significantly
enhanced, relative to its SM value, without conflicting with current
Higgs data.  As the dominant decay mode of the SM Higgs into bottom
quarks is characterized by a rather small coupling, a moderate
enhancement of the charm coupling is sufficient to yield
dramatic changes in the Higgs phenomenology at the LHC. In particular,
we find that current data even allow for the $h\to c\bar c$ mode to
become the dominant Higgs decay channel.  This results in the $h\to
b\bar b$ signal strength being reduced down to $\mathcal{O}(40\%)$
level, which renders observation of this channel rather challenging
for the next LHC run.  However, we argued that a realistic prospective
form of charm-tagging would allow to not only resurrect part of the
lost $b\bar b$ signal but also to obtain signal strengths in the
associated Higgs production channels which are close to the SM expectations by combining
both charm and bottom pairs. 
We also briefly demonstrated that within
several SM extensions an enhanced Higgs to charm coupling can be obtained
through a moderate accidental cancellation between $\mathcal{O}(1)$
couplings of the theory.
 
%%%%%%%%%%%%%%%%%%%%%%%
\section*{Acknowledgements}
%%%%%%%%%%%%%%%%%%%%%%%

We thank Aielet Efrati, Avital Dery and Gavin Salam for helpful discussions. TG is supported by grants from
the Department of Energy Office of Science, the Alfred P. Sloan
Foundation and the Research Corporation for Science Advancement.  The
work of GP is supported by grants from GIF, ISF, Minerva and the
Gruber award.  The seeds of this project were planted in the 25th
Rencontres de~Blois.

%%%%%%%%%%%%%%%%%%%%%%%%%%%%%%%%%%%
\appendix 
%%%%%%%%%%%%%%%%%%%%%%%%%%%%%%%%%%%

%%%%%%%%%%%%%%%%%%%%%%%%%%%%%%%%%%
\section{Unconstrained Higgs fit} \label{append:gen}
%%%%%%%%%%%%%%%%%%%%%%%%%%%%%%%%%%

We report for completeness the results of a global fit to all Higgs data in the most generic case where the eight free parameters $c_V$, $c_{c,b,t}$, $c_\tau$, $c_{gg}$, $c_{\gamma\gamma}$ and $c_{Z\gamma}$ are allowed to deviation from their SM values. We use the freedom of redefining the Higgs boson phase to make $c_V>0$, while the sign of the other parameters remains {\it a priori} unconstrained. However current Higgs data is insensitive to the sign of $c_c$ and we assume $c_c>0$ for simplicity. Following Ref.~\cite{Falkowski:2013dza}, we append the $\chi^2$ function in \Eref{chi2} so as to include EW precision measurements from LEP. LEP measurements are dominantly sensitive to Higgs couplings to weak gauge boson ($c_V$) and photons ($c_{\gamma\gamma}$ and  $c_{Z\gamma}$), which modifies the oblique EW parameters~\cite{PeskinTakeuchi,PomarolStrumia}.
The global fit results are 
\bea
&c_V= 1.04^{+0.04}_{-0.04}\,, \
c_t=0.9^{+1.0}_{-2.7}\, , \
c_c=2.9^{+2.9}_{-2.9}\, ,  \nonumber\\
&c_b=1.26^{+0.33}_{-0.31}\, ,  \
c_\tau=1.19^{+0.25}_{-0.27}\, , \ 
c_{gg}=0.004^{+0.035}_{-0.043}\, ,   \nonumber\\
&c_{\gamma\gamma}=0.0005^{+0.0026}_{-0.0063}\,, \
c_{Z\gamma}=-0.003^{+0.022}_{-0.036} \, . 
\label{fullfit}
\eea
Note that the combination $c_{gg}+1.26\times10^{-2} c_t$, which approximately controls the gluon fusion cross section, is unconstrained by signal strength measurements. In general, when deriving the standard deviations in \Eref{fullfit}, we discarded isolated minima away from SM where large values of $c_{gg}$ cancel against a significantly modified SM top loop contribution in $\sigma_{gg\to h}$ to yield small deviations in Higgs rates.

%%%%%%%%%%%%%%%%%%%%%%%%%%%%%%%%%%
\section{Higgs couplings in GMFV} \label{append:gmfv}
%%%%%%%%%%%%%%%%%%%%%%%%%%%%%%%%%%
In general minimal flavor violation (GMFV) models the Lagrangian relevant to Higgs couplings of the up-type quark sector read~\cite{Kagan:2009bn} 
\beq\label{LGMFV}
	\mathcal{L}_{\rm GMFV} = \mathcal{L}_1 + \mathcal{L}_3\,, 
\eeq
with the marginal operators
\beq
\mathcal{L}_1&=&
	\lambda\big( \bar{Q}_3 \tilde{H}U_3 + \alpha \bar{Q}_a\tilde{H}\chi^a U_3 \nonumber\\
&+&\beta \bar{Q}_3 \tilde{H}\chi^a \phi^{ab}_u  U_b + \gamma \bar{Q}_a\tilde{H}\ \phi^{ab}_u U_b \big) +{\rm h.c.}\,,
\eeq
and the dimension six operators 
\beq
\mathcal{L}_3&=& \frac{g}{\Lambda^2}  H^\dagger H \big(\bar{Q}_3 \tilde{H} U_3 + \kappa \bar{Q}_a \tilde{H}\chi^a U_3\nonumber\\
&+& \eta \bar{Q}_3 \tilde{H} \chi^a \phi^{ab}_u U_b + \zeta \bar{Q}_a \tilde{H} \phi^{ab}_u U_b \big)+{\rm h.c.}\,,
\eeq
where $\alpha, \beta,\gamma,\kappa,\eta$ and $\zeta$ are complex $\mathcal{O}(1)$ numbers. $a,b=1,2$ are first two generation indices, while index $_3$ denotes the third generation. In the mass basis the spurions becomes $\phi_u\propto\left( m_u,m_c \right)$ and $\chi\sim(V_{ts},V_{td})\sim(\lambda_C^3,\lambda_C^2)$, where $V_{ij}$ are CKM matrix elements and $\lambda_C\simeq0.23$ is the sine of the Cabibbo angle. Higher order terms in $\phi_u$ and $\chi$ are neglected.   A similar Lagrangian can be written for the down-type quark sector. 
 The Lagrangian of \Eref{LGMFV} yields the following mass matrix
\begin{align}\label{MuGMFV}
M^u = \frac{\lambda v}{\sqrt 2}
	\begin{pmatrix}
		y_u(\gamma+x \zeta) & 0 & \chi_1(\alpha+\kappa x ) \\
		0 & y_c(\gamma+\zeta x) & \chi_2(\alpha+\kappa x) \\
		y_u \chi_1 (\gamma+\eta x ) & y_c \chi_2 (\gamma+\eta x) & 1+ r x
	\end{pmatrix} \, ,
\end{align}
and Higgs coupling matrix
\begin{align}\label{YuGMFV}
Y^u =  \frac{\lambda}{\sqrt{2}}
	\begin{pmatrix}
		y_u(\gamma+3\zeta x ) & 0 & \chi_1(\alpha+3\kappa x ) \\
		0 & y_c(\gamma+3\zeta x) & \chi_2(\alpha+3\kappa x) \\
		y_u \chi_1 (\gamma+3 \eta x) & y_c \chi_2 (\gamma+3\eta x) & 1+ 3r x 
	\end{pmatrix} \, .
\end{align}
 where we defined $x\equiv v^2/(2\Lambda^2)\,$.

%%%%%%%%%%%%%%%%%%%%%%%%%%%%%%%%%%%%%%%%%

\end{document}